# Need for PRIMA to understand the nature and ISM physical conditions of HST-dark galaxies

**Carlotta Gruppioni**,[a,*] **Lee Armus**,[b] **Matthieu Bethermin**,[c] **Laura Bisigello**,[d] **Denis Burgarella**,[e] **Francesco Calura**,[a] **Ivan Delvecchio**,[a] **Andrea Enia**,[f] **Andreas Faisst**,[b] **Francesca Pozzi**,[a,f] **Giulia Rodighiero**,[d,g] **Alberto Traina**,[a] and **Livia Vallini**[a]

[a]Istituto Nazionale di Astrofisica – Osservatorio di Astrofisica e Scienza dello Spazio, Bologna, Italy
[b]California Institute of Technology, IPAC, Pasadena, California, United States
[c]Université de Strasbourg, CNRS, Observatoire astronomique de Strasbourg, UMR 7550, Strasbourg, France
[d]Istituto Nazionale di Astrofisica – Osservatorio Astronomico di Padova, Padova, Italy
[e]Aix-Marseille University, Laboratoire d'Astrophysique de Marseille, Marseille Cedex 13, France
[f]Università degli Studi di Bologna, Dipartimento di Fisica e Astronomia, Bologna, Italy
[g]Università degli Studi di Padova, Dipartimento di Fisica e Astronomia, Padova, Italy

**ABSTRACT.** One of the main open issues in galaxy formation and evolution is the early assembly of the most massive galaxies and their contribution to the stellar mass and star formation rate densities at early epochs. Massive red sources already in place at $z > 2$ to 3 have been found in deep *Spitzer*-IRAC and ALMA surveys. They are often called optically and near-IR dark, or *HST*-dark, being undetected even in the deepest HST frames. The submillimeter (i.e., ALMA) detection of these sources confirms their high-$z$ dusty nature: they are massive (e.g., $M_* > 10^{10}\,M_\odot$) and dusty star-forming galaxies with estimated redshifts in the 2.5 to 7 range. They seem to lie mostly below the main sequence (MS) of star-forming galaxies and show gas depletion times <1 Gyr. Imaging with the *PRIMA*/PRIMAger instrument over the full 25 to 265 $\mu$m range will allow us to characterize their still uncovered spectral energy distributions between *JWST* and ALMA spectral windows, probing their dust content and properties (e.g., temperature, mass), whereas spectroscopic observations with FIRESS will be the key to investigate the nature of their powering source (e.g., AGN or star formation) and to study the physics of their ISM, by detecting and measuring fine structure lines in the mid- and far-IR domain.

## 1 Introduction

Tracing the history of star formation across cosmic time (or star formation rate density, SFRD) is one of the hottest topics of current galaxy evolution studies. Because stars form in dusty regions, it is instrumental to understand how much star formation (SF) activity in the early Universe is obscured by dust and therefore missed by optical/UV selections, and which objects contribute to this quantity. However, our current knowledge of the SFRD at high-$z$ (>3) is mainly based on galaxies detected in the UV rest-frame[1]), missing the more massive and obscured star-forming galaxies[2] (SFGs). These systems are likely connected to the formation of massive ellipticals, whose existence at $z \sim 2$ to 3 challenges the current galaxy formation theories.[3,4] We are therefore still lacking an important piece in the galaxy formation puzzle: unveiling these missing SFGs, understanding their nature (e.g., do they host an Active Galactic Nucleus, AGN? What is their gas density, metallicity, temperature?), and probing the environment around them is the key to understand massive galaxy formation and evolution and to precisely constrain the total cosmic SFRD at early epochs.

---

*Address all correspondence to Carlotta Gruppioni, carlotta.gruppioni@inaf.it

The bulk of SFGs assemble their mass through a secular process of star formation, following a star formation rate (SFR)–Stellar mass ($M_*$) "sequence" (main-sequence, "MS") up to high redshift,[5–7] and a tight gas–SFR density relation (Kennicutt–Schmidt relation, "KS"[8,9]). Starbursting events (associated with major mergers?[10]) or the cessation of SF ("quenching"[11]) can cause deviations from this equilibrium. Why galaxies suddenly start forming stars at a very high rate, and why they stop, are still open questions. The final products of bursts of SF, followed by a sudden drop of the SFR and subsequent quenching, are the quiescent galaxies observed up to high redshift ($z \simeq 2$ to $3$[4,12,13]). The key to understanding what happened in the galaxy transformation was to capture early massive galaxies during or immediately after their quenching, and investigate their physical conditions.

During the past few years, a significant population of optically undetected galaxies with bright infrared (IR) or submillimetre (sub-mm) emission has been discovered in *Spitzer*/IRAC data,[2,14–16] and some of them with the Atacama large millimeter/sub-millimeter array (ALMA).[17–21] They typically have very red spectral energy distributions (SEDs) and remain undetected even in deep *Hubble Space Telescope* (*HST*) H-band observations [down to $H \geq 27$ mag (AB)], hence their name "HST–dark." Although their SEDs are not well constrained, with a few photometric detections and lack of spectroscopic redshifts, resulting in very large uncertainties on their stellar masses and SFRs,[14,18,22,23] the ALMA-detected ones have better determined SFRs and photometric redshifts (e.g., ALMA detection rules out low-$z$ solutions). The HST-dark galaxies discovered by ALMA represent up to 20% of the total ALMA parent samples[17,20] and likely provide a significant contribution to the high-$z$ SFRD and Stellar mass density (SMD).[2,18,20] In Fig. 1, the *HST*/ACS-I, *Spitzer*/IRAC, and ALMA cutout images are shown for some example HST-dark galaxies (the $3\sigma$ depths of these images are $\sim$0.15 mJy, 29.2 and 25.5 mag for ALMA, HST and IRAC, respectively[24,25]). They show on average higher $M_*$ and shorter gas depletion times (higher star formation efficiencies [SFEs]) than MS galaxies (see Fig. 2; Gruppioni et al. in preparation): can they be the missing progenitors of high-$z$ ellipticals? What are the physical conditions of these early massive systems? What is the mechanism driving their rapid transformation from actively star-forming to quiescent? What is their contribution to the high-$z$ SFRD, and their role in galaxy evolution? However, we must note that the SFRs and $M_{\rm mol}$ estimates for most of the HST-dark galaxies are currently based on a single or very few ALMA detections because no data are available in the far-IR for those galaxies that are too faint to be detected at $>24$ $\mu$m, i.e., by *Spitzer* or *Herschel*. Therefore, a sensitive facility in the far-IR allowing us to identify and characterize these galaxies and derive their main physical properties with much better accuracy than currently possible is indeed necessary.

The new capabilities now offered by the James Webb Space Telescope (JWST[28]) in the near and mid-IR are unique in investigating in unprecedented detail the nature of optical/near-IR-dark (mid-IR bright) galaxies[29,30] in the optical rest-frame. Indeed, with the advent of JWST, new mid-IR bright, near-IR faint galaxies have been identified and started to be characterized,[31–34] confirming the high redshift and massive nature of HST-dark galaxies, their relevance for compiling a complete stellar mass census in the first 1 to 2 Gyr of the lifetime of the universe, and their challenging numbers and properties for the galaxy evolution model, even at redshifts as high as $>10$.[35] Understanding the nature of the HST-dark galaxies is however complex because they have been identified from a variety of selection techniques. There is consensus on the presence of large amounts of dust in those detected by ALMA, although parameters as dust temperature and dust mass are mostly unconstrained, due to the few available far-IR continuum data. This introduces large uncertainties in the derived obscured fraction of star formation in high-$z$ ($>3-4$) galaxies.[36] JWST is characterizing the optical properties of such obscured galaxies and is estimating their stellar masses with much better precision than previously possible, but cannot help constrain their dust and molecular gas properties.

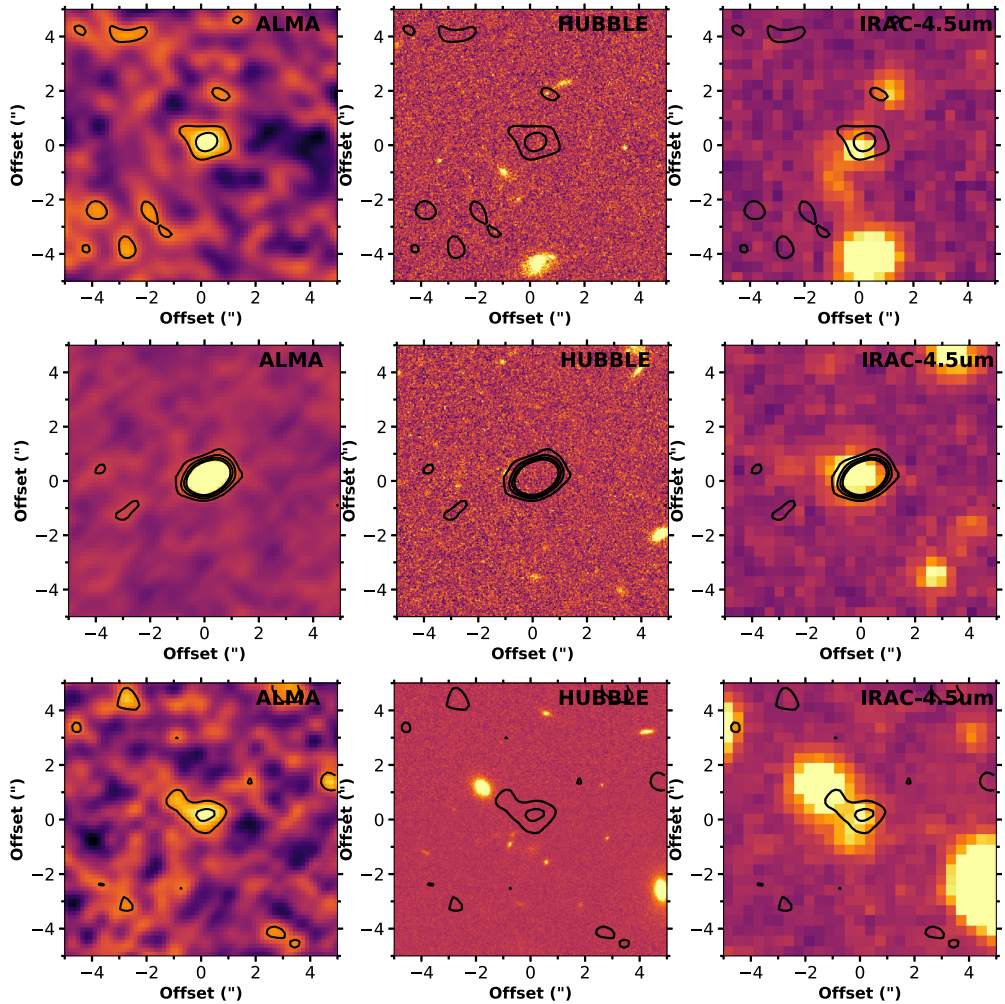

**Fig. 1** Example cutouts of three ALMA-selected HST-dark galaxies: the ALMA-860 $\mu$m (at 3, 5, 7, 9$\sigma$) contours are overplotted in black on ALMA, *HST* ACS-i, and *Spitzer* IRAC-4.5 $\mu$m images. The 3$\sigma$ depths of the images are ~0.15 mJy, 29.2, and 25.5 mag for ALMA, HST, and IRAC, respectively.[24,25] These galaxies are invisible from UV to near-IR and start to shine at mid-IR wavelengths and all the way to sub-mm/mm.

Given the dusty nature of the ALMA-detected, HST-dark galaxies (confirmed by the sub-mm/mm detection) and their red SEDs, the only way to unveil their nature and study the physics of their ISM is through mid-/far-IR photometric and spectroscopic observations. In fact, we have very limited knowledge about the nature of these obscured galaxies, which regulate the SF along the evolution of these galaxies and their metal content at early epochs. Dusty galaxies are the only ones that can tell us the whole story of galaxy and AGN evolution (whose activity peaks at about the same epoch) during that obscured era.

So far, mid- and far-IR facilities such as *ISO*, *Spitzer*, and *Herschel* have allowed us to observe the "obscured" side of the Universe up to Cosmic Noon (i.e., $z \sim 1 - 3$, when galaxy activity peaks) mostly in photometry (by measuring total fluxes of galaxies). Due to sensitivity limits, mid-/far-IR spectroscopy so far has been possible only for nearby galaxy samples,[37] or for very few lensed and hyper-luminous sources at $z > 1$.[38] A sensitive mid- and far-IR photometric and spectroscopic facility will be needed to study the nature and physical properties of statistically sizable samples of obscured galaxies at Cosmic Noon and push the photometric studies of the most luminous dusty sources up to the epoch of reionisation (EoR, i.e., $z \geq 6$).

In this work, we assume a $\Lambda$CDM cosmology with $H_0 = 70$ km s$^{-1}$ Mpc$^{-1}$, $\Omega_\Lambda = 0.7$, and $\Omega_M = 0.3$. Stellar masses and SFRs are normalized to a Chabrier[39] initial mass function (IMF).

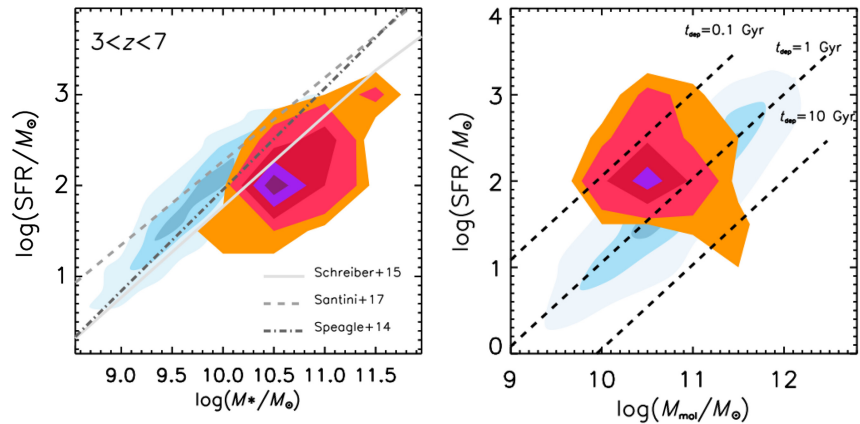

**Fig. 2** Density contours of all the ALMA-detected HST-dark galaxies from the literature (~90 galaxies, orange-to-purple) compared with the locus occupied by the Main Sequence galaxies (blue) in the SFR-$M_*$ (left) and SFR-$M_{mol}$ plane (right): the bulk of the HST-dark galaxies lie below the MS (the different lines show the derivations from different authors: solid,[26] dashed,[27] and dot-dashed[6] lines) and show shorter gas depletion times ($0.1 < \tau_{dep} < 1$ Gyr) than typical MS galaxies ($\tau_{dep} \simeq 1$ Gyr). They are likely massive galaxies on the way to quench their SF, but still forming stars: can they be the progenitors of early massive galaxies?.

## 2 PRobe Far-Infrared Mission for Astrophysics

The *PRobe far-Infrared Mission for Astrophysics* (*PRIMA*) is a far-IR observatory proposed to NASA (recently selected to phase A) equipped with a cryogenically cooled 1.8-m diameter telescope and new-generation detectors (advanced kinetic inductance detectors, KIDs), specifically designed for high-sensitivity imaging and spectroscopic studies in the mid- and far-IR wavelength range. On board of *PRIMA*, there will be two scientific instruments: the *PRIMA* Imaging Instrument (PRIMAger) and the Far-InfraRed Enhanced Survey Spectrometer (FIRESS). PRIMAger is a multiband hyperspectral imager providing 12 independent flux measurements from 25 to 84 $\mu$m across two bands (PHI1: 24 to 45 $\mu$m and PHI2: 45 to 84 $\mu$m, with $R = 10$), and polarimetric capabilities in four broad band filters between 96 and 235 $\mu$m (PPI1 to PPI4, centered at 96, 126, 172, and 235 $\mu$m, with $R \sim 4$). The other instrument, FIRESS, is a versatile, multi-mode survey spectrometer covering wavelengths between 24 and 235 $\mu$m, and offering both low-resolution ($R = 100$) and high-resolution ($R = 12,000$ in Fourier transform spectrometer mode). Thanks to its new-generation detectors cooled down to 100 mK, *PRIMA* will be orders of magnitude more sensitive and faster in scanning large portions of the sky than any previous space missions in the same wavelength range.

## 3 HST-dark Galaxies: A Massive and Dusty Population at High-z

### 3.1 Role of ALMA

In the past few years, ALMA has started revealing serendipitous galaxies in blind surveys and uncovering faint (sub-)mm populations at $z > 2-3$.[17] An important product of these surveys is the discovery of a new class of "dark" galaxies undetected even in the deepest *HST* images (down to $H \geq 27$ mag (AB)[2,17,19,20]). These "HST-dark" galaxies, representing up to 20% of the total ALMA blind samples, are detected in deep *Spitzer*-IRAC images (see Fig. 1). This prominent population of dusty SFGs at $z > 2-3$ likely provides a significant contribution to the high-$z$ SFRD.[2,18,20] In Fig. 3, we show the estimated contribution to the SFRD from HST-dark galaxies from the literature, differently selected (pre-*JWST*, ALMA-selected[18–20]; *H*-dropouts,[2] and radio-selected;[15] by *JWST*[34]). Although the uncertainties in the HST-dark contributions are large (as shown in Fig. 3), due to their small numbers (e.g., typically, we are talking of a few tens in total from all the ALMA blind surveys), different selections (e.g., ALMA, radio, or *H*-dropout) and poorly constrained SEDs, it is clear that the ALMA HST-dark galaxies contribute a significant fraction (>17% to 20%) of the total SFRD at $z > 3$ (at $z \sim 5$ the contribution is similar to the dust-corrected one from all the UV-selected galaxies). This means that the dust-obscured SF is still high up to $z \sim 5-6$, and massive dusty galaxies might play an important role. Their contribution is even more important at the massive end of the stellar mass function (although their mass estimates are very uncertain, being based often on IRAC data only), where they seem to overtake the current volume density estimates of $z > 3$, $\log(M_*/M_\odot) > 10$ galaxies based on rest-frame optical data.[43] The early formation of such large numbers of massive, dusty galaxies is not predicted by current galaxy formation models and simulations[41,44] (see Fig. 3), which underestimate the density of luminous and massive galaxies at $z > 2$ by one to two orders of magnitude.[2,14] The direct implication of such large abundances of massive dusty galaxies at high-$z$ is that our current knowledge of the formation and evolution of massive/luminous galaxies is still far from being complete: the robust estimate of the main physical properties, the ISM conditions, and metallicities of HST-dark galaxies will be crucial for revising the theoretical models considering such a massive and dusty population.

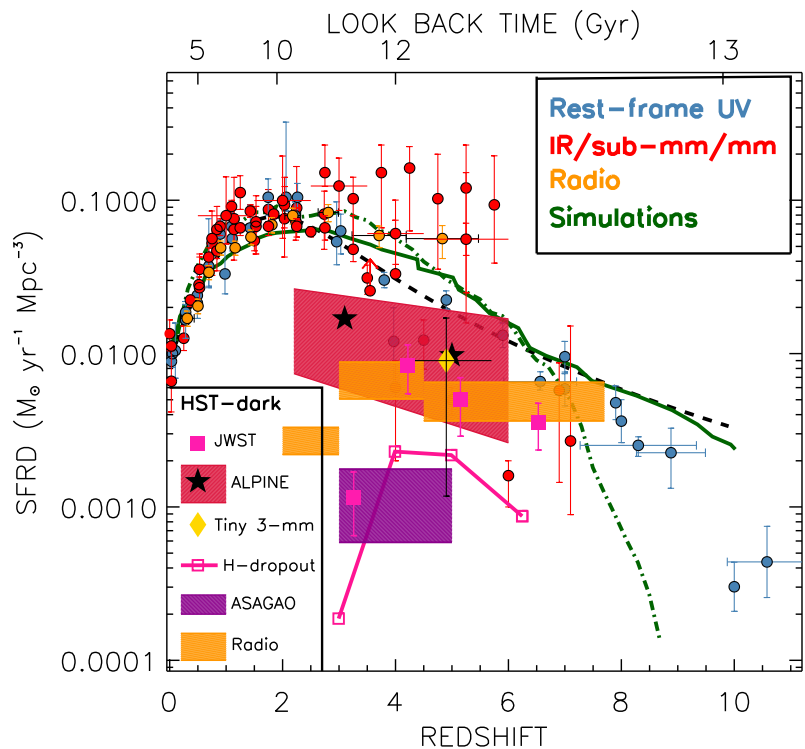

**Fig. 3** Cosmic SFRD as a function of redshift, as derived from IR/sub-mm/mm surveys (red circles), radio (orange circles), and dust-corrected rest-frame UV surveys (blue circles). The black curve is the SFRD by Ref. 40, whereas the green curves are the results of theoretical simulations (solid: Illustris TNG by Ref. 41; dot-dashed: SIDES by Ref. 42). The estimated contribution from different HST-dark galaxy samples is shown with different coloured boxes and symbols, as explained in the legend at the bottom-left corner (Pre-*JWST*: ALMA selected ASAGAO,[19] Tiny 3-mm,[18] and ALPINE[20]; *H*-dropouts[2]; radio-selected[15]; and *JWST*-selected[34]). The contribution of HST-dark galaxies to the cosmic SFRD is significant and becomes more important with increasing redshift, despite the large variations due to different sample selections. Figure readapted from the work by Ref. 20.

ALMA measurements are fundamental for constraining the SF in these early obscured galaxies, because at $z > 3$ the peak of dust emission falls in the sub-mm/mm regime. Moreover, the ALMA continuum data can be used to infer the molecular gas mass under specific assumptions (e.g., that the dust emission in the Rayleigh-Jeans tail of the galaxy SEDs is optically thin and that the low-temperature component of the sub-mm/mm emission contributes to the bulk of the gas mass in massive galaxies ($>10^{10} M_\odot$; Ref. 45). The estimated molecular gas masses of these dusty systems are still significant, and their gas depletion times are short. In Fig. 2, we show the density contours of the ALMA-based SFRs versus $M_{mol}$ and versus $M_*$ (based on the observed near-IR/mid-IR data) for all the ALMA-detected HST-dark galaxies from the literature (see Gruppioni et al. in preparation) and compare them with the same quantities of MS galaxies at similar redshifts. HST-dark galaxies show on average higher $M_*$ and shorter gas depletion times (or higher SFEs) than MS galaxies. The picture emerging from Fig. 2 is that they are still forming stars at a sustained rate (and with a high efficiency), still contain significant molecular gas masses, but have already formed most of their stellar mass: they are likely SF systems on the way to quenching (Fig. 2).

### 3.2 Role of JWST

Recent studies with *JWST* have confirmed previous results based on ALMA and *Spitzer* data, that massive, dusty, high-redshift galaxies were indeed missed from earlier (primarily optically-selected) samples.[29,31,32,35,46] The physical properties of these galaxies were largely unconstrained until the arrival of *JWST*,[29,31–33,35] although they seem to agree well with the previous more uncertain estimates (i.e., HST-dark galaxies are high-$z$ massive systems). *JWST* is probing the rest-frame optical emission of galaxies at $z > 3$, allowing us to obtain more secure photometric redshifts with NIRCam photometry by identifying their Balmer break and to constrain their stellar masses. However, further progress is also possible by using the spectroscopic capabilities of NIRSpec to derive accurate spectroscopic redshifts and probe the physical properties of HST-dark galaxies in the optical rest-frame.[34,47,48] Indeed, the majority of the *JWST* color-selected HST-dark galaxies result to be dusty (estimated attenuation $A_V > 2$ mag), and $H_\alpha$ emitters with equivalent widths indicative of a wide range of recent SF activity, with ∼15% of them being quiescent. All of them (both quiescent and star forming) are confirmed to be early massive systems (though with lower masses with respect to the ALMA + IRAC HST-dark), with stellar masses $\log(M_*/M_\odot) > 9.8$.[29,47,49]

Although *JWST* is crucial in identifying these objects and obtaining spectroscopic and photometric information in the rest-frame optical range, to reach a complete characterization of these very dusty galaxies, of their dust and ISM properties, of the nature of their powering source (i.e., by detecting warm dust heated by an AGN, or AGN lines), we need to fill the wide wavelength gap between *JWST* and ALMA, and observe them in the mid- and far-IR regime, where dust emission peaks.

## 4 Need for *PRIMA*

As mentioned above, to obtain a complete picture of the nature and role of the obscured and elusive population of HST-dark galaxies, we would need to obtain a full characterization of their physical properties, including their ISM (gas and dust). *PRIMA* will be the only facility that will allow us to observe the mid- and far-IR continuum and fine structure lines in dust-obscured galaxies. In particular, *PRIMA* will allow us to unveil the nature of the HST-dark galaxy population by sampling the rest-frame mid- and far-IR part of the spectrum in each source. The 25 to 250 $\mu$m range in $z > 2$ galaxies is crucial to characterize the SED and nature of dusty galaxies, to derive accurate photometric redshifts for those without spectroscopic identification (via the strong PAHs features within the PRIMAges narrow-bands), and to obtain important information on the physical conditions of their dust and gas (through mid- and far-IR fine structure lines). Therefore, photometric and spectroscopic observations of ALMA-selected HST-dark galaxies in this wavelength range with PRIMAger and FIRESS will be the key to investigate the role of such high-$z$ massive and dusty galaxies in early galaxy formation.

As already pointed out in the previous sections, the SEDs of these galaxies, though rising in the mid-/far-IR, are scarcely sampled at wavelengths >5 $\mu$m (see Fig. 5), and their photometric redshifts and physical parameters derived through SED-fitting are largely uncertain. The presence of AGN in the HST-dark population is till debated: although previous studies suggest exceptionally high AGN fraction among ALMA blind surveys (40%[17]), more recent works seem to indicate lower AGN fractions in HST-dark samples.[16] This excess of AGN in the parent samples needs further investigation also among these obscured sources, e.g., by observing them in the range where an AGN, if present, would show up by filling with hot dust the typical dip in host galaxy SEDs (i.e., $\lambda$-rest >2 to 3 $\mu$m[50]). Multi-band mid- and far-IR data and colors are needed to identify and disentangle the AGN from stars. Deriving the AGN occurrence and disentangling their contribution in HST-dark galaxies will be crucial for understanding the role of AGN in the formation of massive galaxies (i.e., are they responsible for quenching SF?) and the formation of super massive black holes (SMBHs) at high-$z$. Given the highly obscured nature of these galaxies, and their non-detection in X-rays, it is possible that the AGN that they might host is also obscured (or Compton thick, CT) or possibly low-luminosity. A recent work by Ref. [51], comparing the expected performances of PRIMA and those of the future X-ray facility NewAthena, showed that even NewAthena will struggle in detecting heavily obscured AGN. In fact, NewAthena proved to be very effective in detecting the most luminous, unobscured, and moderately obscured AGN, whereas PRIMA will be unbearable in identifying low-luminosity and highly obscured AGN, including the CT ones (7500 CT AGN are expected in the planned 1 deg$^2$ PI PRIMA survey). *PRIMA*, through its sensitivity to hot dust and key high-ionization lines over the whole 25 to 250 $\mu$m range is perfectly suited for filling the existing gap in wavelength, crucial to constrain the SEDs of these massive and dusty galaxies and to provide information on their primary powering source (e.g., AGN or SF). In particular, given the range covered by FIRESS and the estimated redshifts of the HST-dark galaxies, it represents the most efficient configuration for detecting [Ne V]$_{14.3,24.3}$ and [O IV]$_{25.9}$ lines, probing the presence of an AGN, and other lines (e.g., [Ne II]$_{12.8}$, [Si II]$_{34.5}$, [O I]$_{45}$, [O III]$_{52}$, [O I]$_{63}$, [O III]$_{88}$), critical for unveiling the physical and chemical properties of the gas inside this obscured population. Therefore, in addition to photometric observations, for the brightest, lower-$z$ subsample of HST-dark galaxies, further investigation into the ISM physics will be possible with FIRESS. In fact, emission-line intensities and emission-line ratios in the mid- and far-IR domain, suffering much lower dust extinction than the optical and UV emission lines, provide unique information on the physical conditions (i.e., electron density and temperature, degree of ionization and excitation, chemical composition) of the gas responsible for emitting the lines within the dust-obscured regions of galaxies with intense SF activity or surrounding an AGN.[52–54] The typical electron density of the emitting gas can be identified through IR fine-structure lines, because different IR fine-structure transitions of the same ion have different critical densities to collisional de-excitation (e.g., [O III] at 52 and 88 $\mu$m). Moreover, the relative strengths of the fine-structure lines of the same element in different ionization stages can probe the primary spectrum of the ionizing source, with these line ratios providing information on the age of the ionizing stellar population and on the strength of the ionization parameter $U$ in single H II regions or in starbursts of short duration. In addition, these line ratios are sensitive to the presence of non-stellar sources of ionizing photons, such as the AGN, with some of these lines with higher ionization potential ([Ne V] at 14.32 and 24.32 $\mu$m, and [O IV] at 25.89 $\mu$m) excited only—or primarily—by an AGN.[55,56] These line diagnostics will enable us not only to identify the primary source of ionization but also to interpret the relative contribution of different ionization mechanisms (AGN/stellar feedback, shocks) to the observed mid-IR line intensities via comparison with state-of-the-art photo-ionisation models.[57] Estimating the AGN contribution in HST-dark galaxies will be necessary for understanding the role of AGN in massive galaxies formation (i.e., are they responsible for quenching SF?) and the early build-up of SMBHs. Fully characterizing the nature of HST-dark galaxies will be crucial to put the still missing piece in the puzzle of galaxy formation and evolution.

### 4.1 Photometric Observations with PRIMAger

We could observe photometrically from 50 to 100 $z > 2$ ALMA HST-dark galaxies (pointed observations on pre-selected sources, minimum possible field of view of $10' \times 10'$ with PRIMger) over the full 25 to 265 $\mu$m wavelength interval, obtaining good statistics over a wide redshift and luminosity range. To estimate the time needed to observe a sizable sample (i.e., 50 to 100) of ALMA-detected HST-dark galaxies, we have considered three luminosity intervals (see Fig. 4):

1. The fainter/higher-$z$ HST-dark galaxies (see the IR luminosity distribution in Fig. 4, and the example SED in the left panel of Fig. 5 showing a $z > 4$, $\log(L_{IR}/L_\odot) \lesssim 12$ HST-dark galaxy from the ALPINE serendipitous sample[20]) for which we estimate a >5$\sigma$ detection in about 10 h at $\lambda > 30$ $\mu$m (more time at shorter wavelength, and in about 1 h time longward of 90 $\mu$m). Observing these galaxies will be challenging, and we could observe a few of them, some possibly within the same field of view, to keep a reasonable observing time.

2. The bulk of the HST-dark population ($3 \lesssim z < 4$, $\log(L_{IR}/L_\odot) \geq 12.5$, for which we estimate to reach a 5$\sigma$ detection in about 1 h. To estimate the exposure time for the bulk of galaxies, we have considered the median SED obtained for all the HST-dark galaxies from ALPINE,[20] AGS,[17] ALMA H-dropout,[2] and ASAGAO.[19]

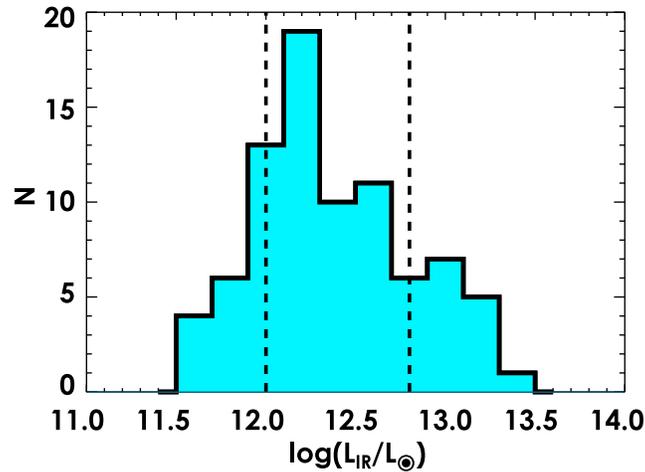

**Fig. 4** Total IR luminosity distribution for all the ALMA-selected/detected HST-dark galaxies from the literature (Gruppioni et al. in preparation). The two vertical dashed lines show the three luminosity regions considered to estimate the observing times with *PRIMA*.

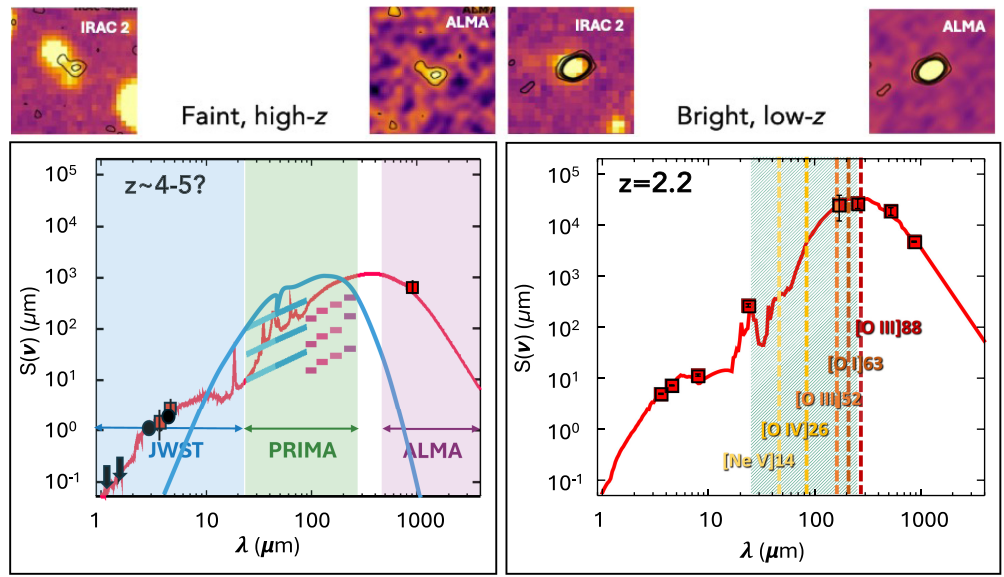

**Fig. 5** Left: Example SED of one of the faint (e.g., $L_{IR} \lesssim 10^{12} L_\odot$, $z > 4$) ALMA-detected HST-dark galaxies (pre-*JWST* detections, i.e., ALMA + IRAC data, are shown by red filled squares, whereas the *JWST* fluxes and upper limits are shown by black circles and downwards arrows, respectively). The range that will be covered by *PRIMA* will contain either PAH features or warm dust heated by an AGN. Note that the AGN component has been added for demonstration purposes only, it is not part of the fit, as the range is totally unconstrained without PRIMA. The black diagonal and horizontal colored lines show the average $5\sigma$ sensitivities of PRIMAger PPI and PPH bands, respectively, in 1 h (upper), 10 h (middle), and 100 h (lower). Right: Example SED of one of the brightest and lower-redshift ($L_{IR} \simeq 10^{12.8} L_\odot$, $z_{phot} \simeq 2.2$) ALMA HST-dark galaxy (detected also by *Herschel* and *Spitzer*/MIPS[20]): the observed data are shown as red filled squares, whereas the red curve shows the best-fit SED. The wavelength range that will be covered by *PRIMA* is shown as a green area. The location of the main IR lines is shown as vertical dashed lines. At the top of each figure, we show the *Spitzer*-IRAC and ALMA images used to first identify and characterise these sources as HST-dark. *PRIMA* covers a critical part of the SED in dusty galaxies at these epochs and will be extremely valuable for measuring PAH features, identifying hot dust heated by an AGN, and detecting strong emission from IR cooling lines in this elusive population.

3. The brighter/lower-$z$ (see Fig. 4, and the example SED shown in the right panel of Fig. 5: $z < 3$, $\log(L_{IR}/L_\odot) \gtrsim 12.8$, from the ALPINE serendipitous sample[20]) for which we estimate to reach $5\sigma$ in less than 0.1 h over the whole 24 to 265 $\mu$m wavelength range.

The time estimates have been obtained by rescaling the sensitivities reported in the PRIMAger factsheet according to the observing area (provided in the PRIMA webpage).[58] Because the full range is observed simultaneously and shorter times are needed to reach the expected fluxes (or confusion) in the PRIMAger PPI bands (i.e., ~90 to 265 $\mu$m), an average estimate of 30 min$^{-1}$ h per source will likely meet our goal. We would therefore estimate that in about 50 h, we will be able to observe between ~50 and 100 HST-dark galaxies with PRIMAger (because they are pointed observations, in our calculations, we have considered the smallest possible field of view that corresponds to $\sim 10' \times 10'$ observations centered on each target, allowing simultaneous coverage of $4' \times 4'$ with both PHI and PPI; L. Ciesla, private communication). Note that the estimated times aim to provide only an order of magnitude of the time needed to observe a sizable sample of HST-dark galaxies with PRIMA. At the time of observation, depending on the exact fluxes of the sources and on the balance between the number of fainter (requiring long observing times) and brighter galaxies (requiring much shorter times), we will set the exact number and luminosity distribution of the targets. Our time estimates are all conservative because we do not consider the presence of an AGN: if warm dust heated by the AGN fills the galaxy dip, it will fall in the PRIMAger bands, allowing us to detect our targets in a significantly shorter time (see Fig. 5).

In Fig. 6, we show how the precision in the measurement of $L_{IR}$ will improve by including the PRIMAger bands, compared with what we can obtain now for the ALMA HST-dark galaxies with only IRAC + ALMA and rarely also MIPS and *Herschel* detections. In the figure, we show the normalized distributions of the difference between the estimated and the true (logaritmic) values of $L_{IR}$ obtained by simulating and fitting galaxy SEDs including PRIMAger flux measurements, as obtained by Ref. 59 (from purple to violet-red: PHI only, PHI+PPI1, PHI + PPI1 + PPI2, respectively). We compare these distributions with the normalized distribution of the $L_{IR}$ uncertainties obtained by fitting the current SEDs of the ALMA HST-dark galaxies (without PRIMAger fluxes: cyan curve, based on most cases on a single (or a few) ALMA detection(s) + IRAC data only). As shown in the legend, for the PRIMAger simulated results, the lines are color-coded according to the set of filters used in the fitting (for details on the model parameters, we refer to Ref. 59). The dashed lines show the same distributions with the PHI + PPI1 + PPI2 set of filters but assuming a SN of 5 instead of 10 in PHI1_4, whereas the dotted line shows the results using the confusion noise for the deep catalogue from Ref. 60, obtainable in regions where ancillary information exists (e.g., from the Roman Space Telescope). The improvement in the precision of $L_{IR}$ that we can achieve by introducing the PRIMAger measurements is significant, passing from a width of the distribution of $\simeq 0.5$ to $\simeq 0.1$ to 0.15.

### 4.2 Spectroscopic Observation with FIRESS

We estimate to be able to observe spectroscopically the ~20 to 25 brightest (in $L_{IR}$, e.g., $>10^{12.8} L_\odot$ at $z < 3$) ALMA HST-dark galaxies with FIRESS in low-resolution mode ($R \simeq 100$). These brighter galaxies constitute the lower-$z$ tail of the HST-dark population distribution (typically ranging from $z = 2$ to $z = 6$ to 7 and peaking at $z \simeq 3.5$): given the nominal sensitivities of FIRESS, this lower-$z$ subsample is the only one observable in a reasonable— though significant—amount of time. Based on the PRIMA ETC (assuming a line sensitivity $S_{line}(5\sigma, 1\ h) \sim 2 \times 10^{-19}$ W/m$^2$), by considering the local relations between $L_{line}$ and $L_{IR}$ by Ref. 61 for $L_{IR} = 10^{12.8} L_\odot$ and $z = 2.2$ (i.e., the properties of the galaxy shown in the right panel of Fig. 5, corresponding to average values for the bulk of the brightest HST-dark galaxies, whose IR luminosity distribution is shown in Fig. 4), we estimate that with exposure times of ~2 h/source we will be able to detect [Ne II]$_{12.8}$, [Si II]$_{34.5}$, [O I]$_{63}$, [O III]$_{88}$, and [O IV]$_{25.9}$ (by considering for the latter the relation provided for a low AGN fraction, $f_{AGN} < 40\%$ the 5 to 40 $\mu$m range, valid also for no or negligible AGN). The [Ne V]$_{14.3,24.3}$ lines instead will be detectable in a reasonable time (about 1.6 to 2 h) only for higher AGN fractions, i.e., $f_{AGN} > 40\%$. If the fraction of AGN will be >40% over the 5 to 40 $\mu$m range, we estimate that the [O IV]$_{25.9}$ will be observable in only 0.4 to 0.5 h. We would therefore estimate that in about 100 h, we will be able to detect the stronger IR lines in ∼25 HST-dark galaxies at $z \simeq 2$ to 2.5 over the full FIRESS spectral range (two pointings per source to cover the four bands). Note that the time estimates are conservative, because the times can be shorter for some lines if we use ratios between $L_{line}$ and $L_{IR}$ from different works (e.g., [Ne II]$_{12.8}$, the SFR tracer, can be observed in only 10 min if we consider the relation of Ref. 62). If we take the line luminosities obtained for $L_{IR} = 10^{12.8} L_\odot$ through the Ref. 61 relations as limiting values, we can give a rough estimate of the limiting SFR and BHAR (if an AGN is present) that we can reach by integrating 2 h with FIRESS. From the [Ne II] luminosity, we obtain the IR luminosity due to SF (using the relation between $L_{line}$ and $L_{IR}^{SF}$), then we convert it to SFR using the Ref. 9 relation for a Ref. 39 IMF, obtaining $SFR_{lim} \simeq 300 \, M_\odot \mathrm{yr}^{-1}$. From the [O IV] luminosity (obtained considering $f_{AGN} < 40\%$), we first derive the bolometric AGN luminosity, $L_{BOL}^{AGN}$, through the Ref. 61 relation between $L_{line}$ and $L_{BOL}^{AGN}$; then, we obtain the limiting accretion onto the BH by considering the relation between $L_{BOL}^{AGN}$ and the BHAR, $\dot{M}_{BH}$[63]:

$$L_{BOL}^{AGN} = \frac{\dot{M}_{BH}(1-\epsilon)}{c^2 \epsilon}, \quad (1)$$

where $c$ is the speed of light and $\epsilon$ is the accretion efficiency (commonly assumed $\epsilon = 0.1$). By making the opportune conversions, we obtain a $(\dot{M}_{BH})_{lim} \simeq 5 \, M_\odot \mathrm{yr}^{-1}$ for $L_{BOL}^{AGN} = 10^{46.5}$ erg/s, values in the range covered by type 1 AGN or QSOs.

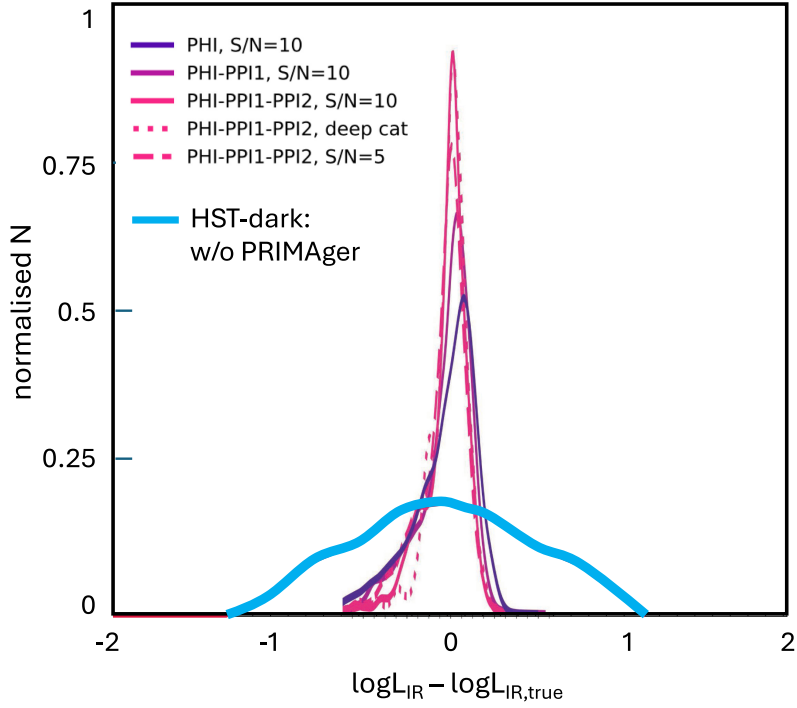

**Fig. 6** Normalized distributions of the difference between the estimated and the true values of $L_{IR}$ obtained by simulating and fitting SEDs with PRIMAger flux measurements (from purple to violet-red: PHI only, PHI + PPI1, PHI + PPI1 + PPI2[59]), compared with the distribution $L_{IR}$ uncertainties obtained for the ALMA HST-dark galaxies without PRIMAger fluxes (cyan curve, based on most cases on a single or a few ALMA + IRAC detections only). As shown in the legend, for the PRIMAger simulated results, the lines are color-coded according to the set of filters used in the fitting (for details on the model parameters, we refer to Ref. 59). The dashed lines show the same distributions, with the PHI + PPI1 + PPI2 set of filters but assuming a SN of 5 instead of 10 in PHI1_4, whereas the dotted line shows the results using the confusion noise for the deep catalog from Ref. 60, obtainable in regions where ancillary information exists (e.g., from the Roman Space Telescope).

This ambitious experiment, expensive in terms of observing time, will be unique in unveiling the physics of the high-$z$ massive dust obscured population whose existence challenges the current galaxy formation and evolution models and is possibly in tension with the $\Lambda$CDM standard cosmology.[30,34,49] A comprehensive characterization of this population is impossible to obtain by relying only on facilities sampling shorter wavelengths (e.g., JWST).

## Disclosures

The authors declare there are no financial interests, commercial affiliations, or other potential conflicts of interest that have influenced the objectivity of this research or the writing of this paper.

## Code and Data Availability

The data used in this study were obtained from the literature, in particular from the following surveys: ALMA GOODS-S,[17] ASAGAO,[19] H-Dropouts,[2] ALPINE,[20] and AS2UDS.[21]

## Acknowledgments

CG and FC acknowledge the support by the Italian MIUR Progetti di Ricerca di Rilevante Interesse Nazionale (PRIN) (Grant Nos. 2017–20173ML3WW_001 and 2022_n._20229YBSAN).

**Carlotta Gruppioni** is a senior staff researcher at the Italian National Institute of Astrophysics (INAF), Osservatorio di Astrofisica e Scienza dello Spazio di Bologna. She received her master's degree in physics and her PhD in astronomy from the University of Bologna in 1993 and 1987, respectively. She is a leading expert in infrared extragalactic astrophysics, galaxy, and AGN formation and evolution. She has long-term experience with mid- and far-IR satellite surveys. She has extensively worked on the scientific analysis of *ISO*, *Spitzer*, and *Herschel* survey data and on the study for the ESA/JAXA *SPICA* mission configuration and exploitation. She is currently a Co-I of *PRIMA*. She has been actively involved in the Herschel GTO PACS extragalactic survey PEP and in the ALMA ALPINE large programme, leading the luminosity function studies.

Biographies of the other authors are not available.